\documentclass[reprint,prl,doublecolumn, amsmath,amssymb]{revtex4-1}

\usepackage{xcolor}
\usepackage{graphicx}
\usepackage{dcolumn}
\usepackage{bm}
\usepackage[colorlinks=true,citecolor=blue]{hyperref}

\begin{document}

\preprint{APS/123-QED}

\title{Tunable bimodal explorations of space from memory-driven deterministic dynamics}

\author{Maxime Hubert}
\affiliation{GRASP, Institute of Physics, Universit\'{e} de Li\`{e}ge, 4000 Li\`{e}ge, Belgium, EU}
\affiliation{Current address: PULS Group, Institut f\"ur Theoretische Physik, Friedrich-Alexander-Universit\"at Erlangen-N\"urnberg, 91058 Erlangen, Germany, EU}

\author{St\'{e}phane Perrard}
\affiliation{Laboratoire de Physique de l'ENS, CNRS UMR 8550 ENS and PSL University, 75005 Paris}

\author{Matthieu Labousse}
\email{matthieu.labousse@espci.psl.espci.fr}
\affiliation{Gulliver, CNRS UMR 7083, ESPCI Paris and PSL University, 75005 Paris France, EU}

\author{Nicolas Vandewalle}
\affiliation{ GRASP, Institute of Physics, Universit\'{e} de Li\`{e}ge, 4000 Li\`{e}ge, Belgium, EU}

\author{Yves Couder}
\affiliation{Mati\`{e}re et Syst\`{e}mes Complexes, CNRS UMR 7057, Universit\'{e} Paris Diderot, Sorbonne Paris Cit\'{e}, 75013 Paris, France, EU}

\date{\today}

\begin{abstract}
We present a wave-memory driven system that exhibits intermittent switching between two  propulsion modes in free space. The model is based on a point-like particle emitting periodically cylindrical standing waves. Submitted to a force related to the local wavefield gradient, the particle is propelled, while the wave field stores positional information on the particle trajectory. For long memory, the linear motion is unstable and we observe erratic switches between two propulsive modes : linear motion and diffusive motion. We show that the bimodal propulsion and the  stochastic aspect of the dynamics at long time are generated by a Shil'nikov chaos. The memory of the system controls the fraction of time spent in each phase. The resulting bimodal dynamics shows analogies with intermittent search strategies usually observed in living systems of much higher complexity. 
\end{abstract}

\maketitle 
	At the individual level, random switches between straight line motion and erratic changes in direction has been observed in many living systems such as microscopic bacteria~\cite{Berg1972}, flies~\cite{Strauss_1998,Neuser2008} or macroscopic animal foraging~\cite{Bergman_2000}. The theoretical description of animal foraging based on bimodal exploration of space probably cannot be separated from a fine analysis of the spatial heterogeneities of their environment~\cite{Wiens_1976,Senft_1987}. However, in some specific cases, the origin of bimodal motions may be justified using optimal search strategies in the absence of cues~\cite{Benichou2011}. Indeed, without any cues about the target location, bi-modal motions can be encountered when exploration and exploitation tasks are not performed simultaneously, for instance in fishes looking for prey of various sizes~\cite{Obrien_1989}. In contrast, when the searcher can perform both tasks simultaneously, Levy flights strategies~\cite{Viswanathan99,Reynolds09} with exponents depending on the target properties are optimal even if their observations and relevances in the context of animal motions has been raised~\cite{Benhamou2007,Pyke_2014,Edwards2007}. 
  	
  	The theoretical description of these individual trajectories usually involves some stochasticity: from the pioneer work of Pearson~\cite{Pearson1905} to the numerous recent theoretical analysis using persistent random walks~\cite{Bartumeus_2005,Book_foraging2011}. Apart from such stochastic models individual trajectories of erratic aspects may also be obtained from deterministic rules, in which stochasticity becomes an emergent behaviour. Emergent complexity from simple deterministic model rules would provide a robust artificial implementation of bimodal explorations. The statistics of both phases of motion would not be pre-set in an arbitrary manner. Instead it could be adjusted by changing one single tunable parameter. In this Letter, we consider a single point-like particle ruled by a deterministic set of equations, which exhibits two intertwined modes of space exploration : ballistic and local erratic motions. The model is inspired from chaotic systems, time-delayed differential equations, and hydrodynamics experiments.\\

	Chaotic dynamics provide a first source of inspiration. Intermittent switches between laminar and erratic phases in dynamical systems were first found experimentally in thermal convection~\cite{Berge1980}. Their origin has been rationalized using low-dimensional nonlinear models~\cite{Berge_1984,Manneville}. However, the low dimensionality of those models prevents the observation of diffusive dynamics. Nevertheless with a deterministic set of equations, diffusive behaviours with kinks can be observed using Wolfram cellular automata~\cite{Grassberger1984} or \emph{Red queen walk}~\cite{Freund1992}. In these cases, information is encoded into the environment and shows that memory-driven dynamics may exhibit emergent diffusive properties.
	
	As a consequence, delayed differential equations provide a second natural source of inspiration. Adding a delay in differential equation~\cite{Erneux2009,Erneux2017,NATObook} can trigger instabilities and intermittency whose applications range from car following model~\cite{Reuschel50,Pipes1953,Helbing2001}, complex population dynamics variations in trophic levels~\cite{May1973}, instabilities in delayed logistic dynamics~\cite{Buchner2000} to laser destabilization~\cite{Mandel1987}. In contrast, stabilizing effects of time-delayed term is also used in control theory, in stochastic dynamics~\cite{Hiroyasu2017} or phototactic robots~\cite{Mijalkov2016,Leyman2018}. Dynamics with multiple delays have also been investigated and the nonlocality in time offers a hierarchy of mathematical complexity that may serve as computational principles, for example trough spatiotemporal spikes coding in the context of deterministic neural networks~\cite{Popovych2011} and that may be used to generate bimodal distribution of motion. The addition of long lasted memory in numerical stochastic processes such as in Elephant Random Walk~\cite{Schutz2004} can also trigger different phases of motion~\cite{Silva2008}. Remarkably a single tunable non-Markovian model~\cite{Kumar2010} exhibits diffusion, subdiffusion, superdiffusion or signature of intermittent behaviors.
	
	An hydrodynamics experiment provides the third source of inspiration and a new conceptual framework to store information.  The implementation of a spatially extended and tunable memory in a simple deterministic physical system was achieved with walking droplets on a vertically vibrated bath~\cite{Walker_Nature,Bush2014}. A silicone droplet compelled to bounce on a vertically vibrated liquid surface generates by its successive impacts monochromatic cylindrical standing waves, thanks to the proximity of the Faraday instability. In return, the waves propel the droplet along the surface. The waves are slowly damped in time, with the temporal decay controlled by the bath acceleration~\cite{Eddi2011}. Besides this particular experimental implementation, it is an example of physical trajectory encoding in a surrounding wave medium~\cite{Eddi2009}. The center of each circular wave pattern is located at the exact previous position of the walker such that a positional information is stored in an oscillating wave field~\cite{Eddi2011,Oza2013,Milewski2015}. While deposition of information along a path like for ants may lead to bio-inspired algorithmic principles~\cite{Bonabeau2000}, the wave persistence defines a memory time during which the positional information is stored in a way suitable for defining a Turing machine~\cite{Perrard2016}. Besides wave-particle inspired dynamics~\cite{Walker_Nature,Protiere_JFM,Eddi2009,Nachbin_2017,Bush2014,Fort_PNAS,Molacek2013a,Molacek2013b,Bush2014,Oza2013,Oza2014_JFM,Perrard2014_Nature,Labousse2014_NJP,Filoux_2015,Labousse2016,Gilet_2016,Milewski2015,Durey_2017,Hubert_2017,Saenz_2018}, walking droplets exhibit cascades of bifurcation to chaos in Coriolis and Coulomb force field~\cite{Tambasco2016} as well as intermittency in harmonic potential~\cite{Perrard2014_PRL,Perrard2018,Tambasco2016,Labousse2014_NJP,Perrard_2014_these,Labousse_2014_these,Durey_2018}. Non steady propulsions have been reported in asynchronous bouncing modes~\cite{Oistein_2013,Sampara2016} and speed limit cycle and chaotic behaviour for the free particle~\cite{Bacot2018} has been investigated for synchronous bouncing modes. In this article we leverage the wave-memory to implement three modes of motion at the single particle level in a same model: ballistic, diffusive and intermittent motions. 

	The experiment-inspired numerical model is implemented as follow. The iterative dynamics consists in the parallel equations of motion of a particle at the position $\vec{r}_k$ at the $k$th bounce and its associated wave field $\zeta(\vec{r},t)$. The motion is decomposed in two phases of respective duration $t_1$ and $t_2$ such that $t_1+t_2 = T$ is the wave period. Phase $1$ corresponds to a free flight motion, in which the particle follows a planar motion above the wave field at constant horizontal velocity. Phase $2$ corresponds to the contact with the surface which yields both interaction with the waves and energy dissipation. The particle slides on the surface for a duration $t_2$ with an exponentially decaying speed before taking off again. At the peculiar instant between phase 1 and phase 2, the particle gets a kick of momentum proportional to the local slope of the wave field. The wave field is updated simultaneously. A new standing cylindrical wave is added, centred at the current position of the particle. The total wave field $\zeta(\vec{r},t)$ after the $N$-th bounce at time $t_N$ writes
\begin{equation}
	\zeta(\vec{r},t_N)= \zeta_0 \sum_{n=1}^N J_0\left(\frac{2\pi}{\lambda}\vert\vec{r}-\vec{r}_n \vert\right)e^{-\frac{\vert\vec{r}-\vec{r}_n \vert}{\delta}}e^{-\frac{t_N-t_n}{\mathrm{Me}}}
\label{wavefield}
\end{equation} 
where $\zeta_0$ accounts for the amplitude of one standing cylindrical Bessel wave $J_0$ and $\lambda$ is the wavelength. In the inspiring experiments $\lambda\simeq 5$ mm and is used to normalize the lenght scale.  The memory parameter $\mathrm{Me}$ sets the effective number of active wave sources, and $\delta = 2.5 \lambda$ accounts for the spatial attenuation of viscous dissipation~\cite{Eddi2011} (see Supplemental Materials for the numerical values and appendix A of~\cite{Hubert_2018} for a detailed description of the algorithm). The control parameter $\mathrm{Me}$ is expected to play a key role in the dynamics. \\

\begin{figure}[h!]
	\centering
	\includegraphics[width=\linewidth]{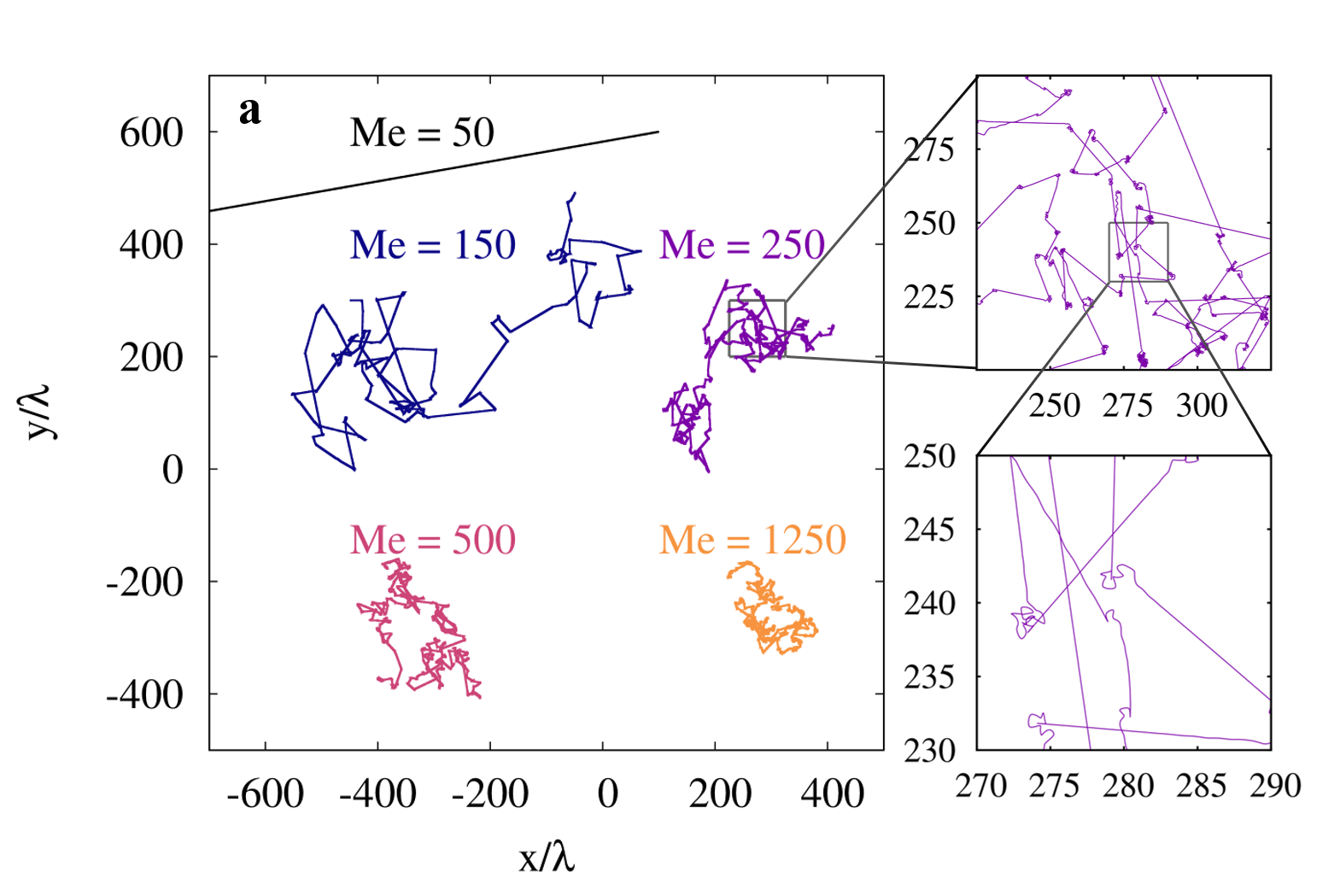}
	\includegraphics[width=0.9\linewidth]{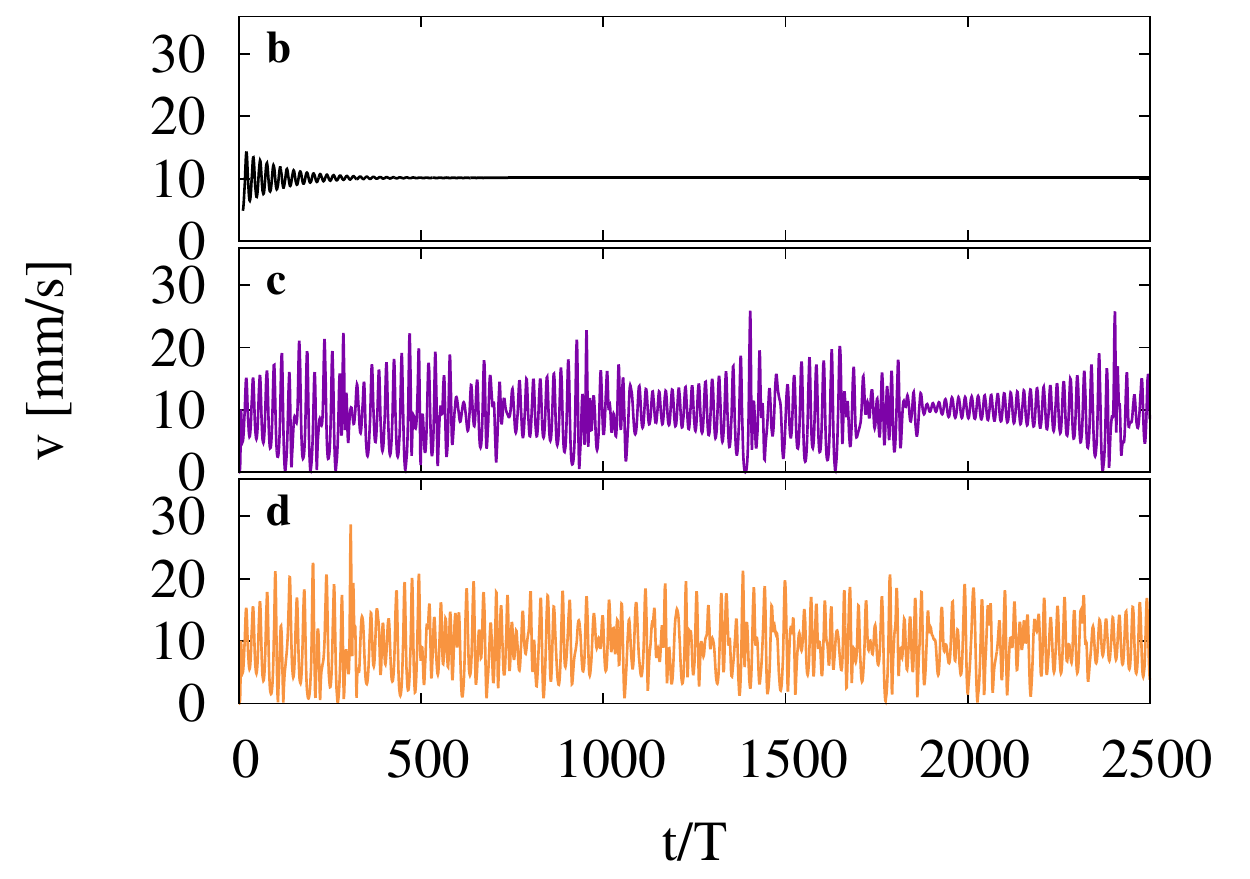}
	\caption{(color online) Evolution of the wave memory-driven particle dynamics with increasing memory (a) Particle trajectories obtained for increasing values of $\mathrm{Me} = 50, 150, 250, 500, 1250$. They correspond to the same simulation time. Zoom into one trajectory details shows evidences of a bimodal dynamics. Straight lines alternate with jiggling motion. (b)-(d) Particle speed as a function of time for increasing memory parameters (b) $\mathrm{Me}=50$, (c) $\mathrm{Me}=250$, (d) $\mathrm{Me}=1250$. Above $\mathrm{Me}=140$, the dynamics is unstable. Duration of the laminar phases decrease with $\mathrm{Me}$.}
	\label{fig:Fig1}
\end{figure}

	The particle trajectories for increasing memory parameter $\mathrm{Me}$ are shown in Fig.~\ref{fig:Fig1}(a) (see also the supplemental movies SM1 and 2). They are obtained from the same initial conditions and a simulation time of $\Delta T_{\mathrm{simu}} = 2.5\,10^5 T$. For $\mathrm{Me} = 50$, the particle moves along a straight line, as described in previous works~\cite{Oza2013}. This linear motion results from a stable balance between propulsion from the wavefield and dissipation by wave emission and viscous friction~\cite{Protiere_JFM}. From $\mathrm{Me} = 150$, straight line motions are interspersed by sudden changes of direction. Successive zooms in Fig.~\ref{fig:Fig1}(a) reveal erratic changes of direction at the scale of the wavefield wavelength, alternating with periods of linear motion. For larger $\mathrm{Me}$ the length of straight line trajectories shorten on average : the memory parameter affects the rates of switching dynamics between both modes of space exploration. The switch between linear and erratic motion finds a signature in the particle speed. Figure~\ref{fig:Fig1}(b,c,d) presents the temporal evolution of the velocity modulus $\mathrm{v}$ for increasing memory corresponding to the trajectories shown in Fig.~\ref{fig:Fig1}. For $\mathrm{Me} = 50$ (Fig.~\ref{fig:Fig1}b), transient speed oscillations decrease exponentially in time, and the dynamics converges toward an uniform linear motion. Note that a similar behavior has been observed both experimentally and numerically by Bacot {\it et al.}~\cite{Bacot2018}. For $\mathrm{Me} = 150$ (Fig.~\ref{fig:Fig1} (c)), the temporal signal of the particle speed is a succession of oscillations of slowly increasing amplitude and more complex excursions of short duration. On a longer time scale, we observe intermittent switches between slow diverging speed oscillations (laminar phase) followed by erratic motion (chaotic phase). For longer memory parameter, the laminar phases shorten while erratic phases become predominant (Fig~\ref{fig:Fig1}(b,c,d)). 
	
\begin{figure}[tbhp]
	\centering
	\includegraphics[width=0.9\linewidth]{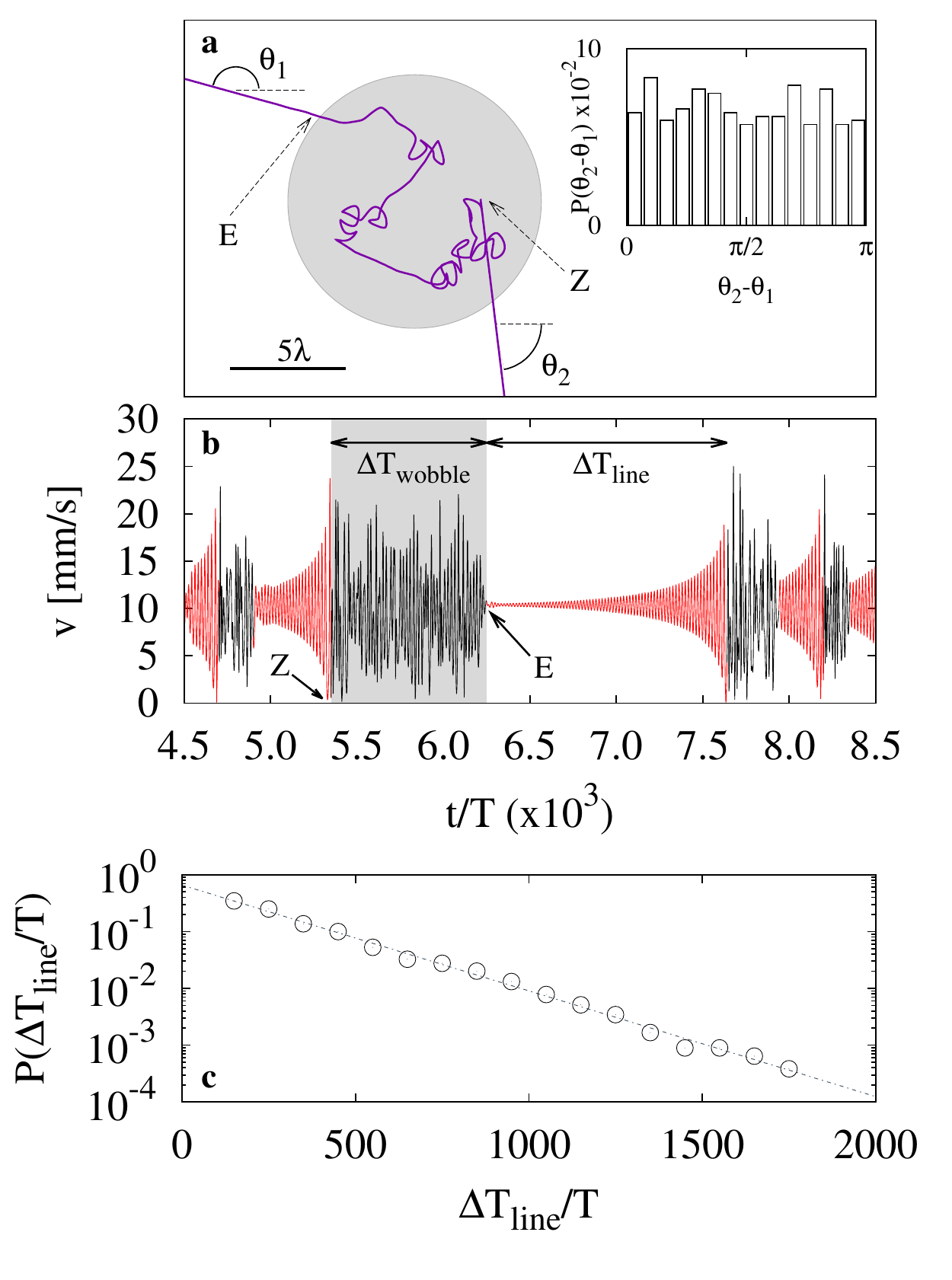}
	\caption{(color online) Description of the bimodal dynamics (a) Zoom on the path of one wobble phase. \textit{Inset}: histogram of $\theta_2 - \theta_1$ showing a uniform distribution. (b) Speed as a function of time. The gray zone corresponds to the tumble phase shown in Fig.2a. Wobbling motion corresponds to erratic fluctuation of speed (chaotic phase) while straight line motion corresponds to slow diverging oscillation of speed (laminar phase). (c) Probability distribution (lin-log.) of the time spent in straight line for $\mathrm{Me}=250$. }
	\label{fig:Fig2}
\end{figure}

The relation between the trajectories and the speed is described in Fig.~\ref{fig:Fig2}. Figure~\ref{fig:Fig2}(a) zooms on a wobbling phase between two rectilinear parts of the dynamics at $\mathrm{Me} = 250$. Figure~\ref{fig:Fig2}(b) shows the associated speed oscillations. Linear motion coincides with speed oscillations, while erratic trajectories coincides with erratic speed fluctuations. We use a local radius of curvature of the trajectory greater than $\lambda_F$ for at least 100 bounces as a criterion for defining a straight line. Starting the description from the beginning of a laminar phase, the divergence of speed oscillations leads to vanishing velocity (Zero (Z) point in Figs~\ref{fig:Fig2}a and ~\ref{fig:Fig2}b ). This specific moment corresponds to a sharp change of direction in the particle trajectory. At Z, the particle hits the surface wavefield with a positive slope which triggers a back motion and initiates the transition from the linear phase to the wobbling phase. Then the particle navigates erratically in a confined region of space during a period $\Delta T_{\mathrm{wobble}}$. The complex trajectory in the wobbling phase is attributed to the dynamical trap of the particle by the wave field structure. After a certain period of time the wobbling phase ceases and the particle enters again in a phase of straight line motion of duration $\Delta T_{\mathrm{line}}$ (E point in Fig.~\ref{fig:Fig2}(a),(b)). The spatial extent of the trajectory in the wobbling phase is of the order of ten wavelengths as emphasized in grey of Fig.~\ref{fig:Fig2}(a). The histogram inset of Fig.~\ref{fig:Fig2}(a) reveals that the erratic phases perform a statistically isotropic reorientation of the trajectory. The duration of straight line motions follows an exponential  distribution as evidenced in Fig.~\ref{fig:Fig2}(c). As the memory parameter is increased, the distribution of time in the laminar phase remains exponential and the average time decreases.

\begin{figure}[tbhp]
	\centering
	\includegraphics[width=0.9\linewidth]{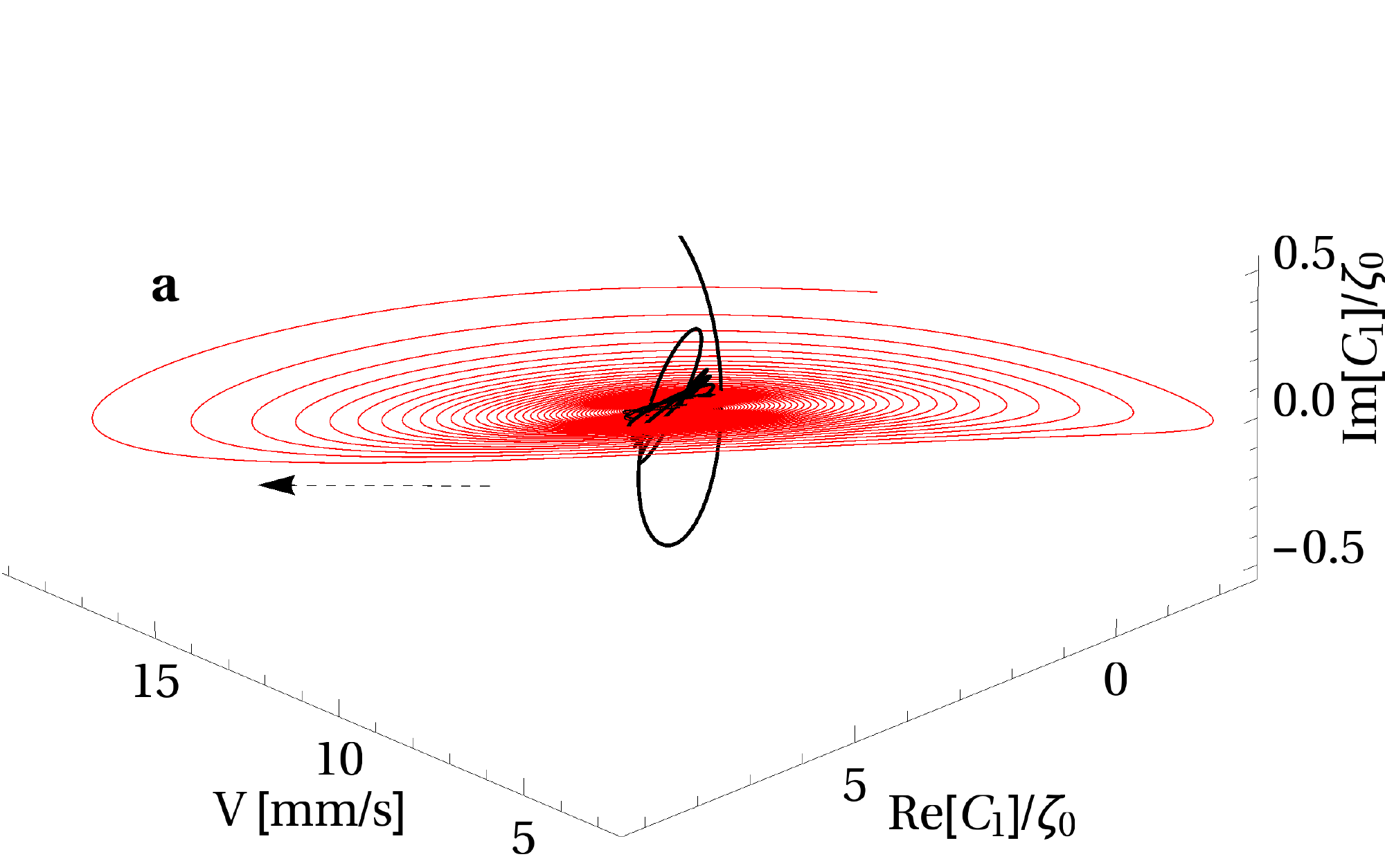}
	\includegraphics[width=0.9\linewidth]{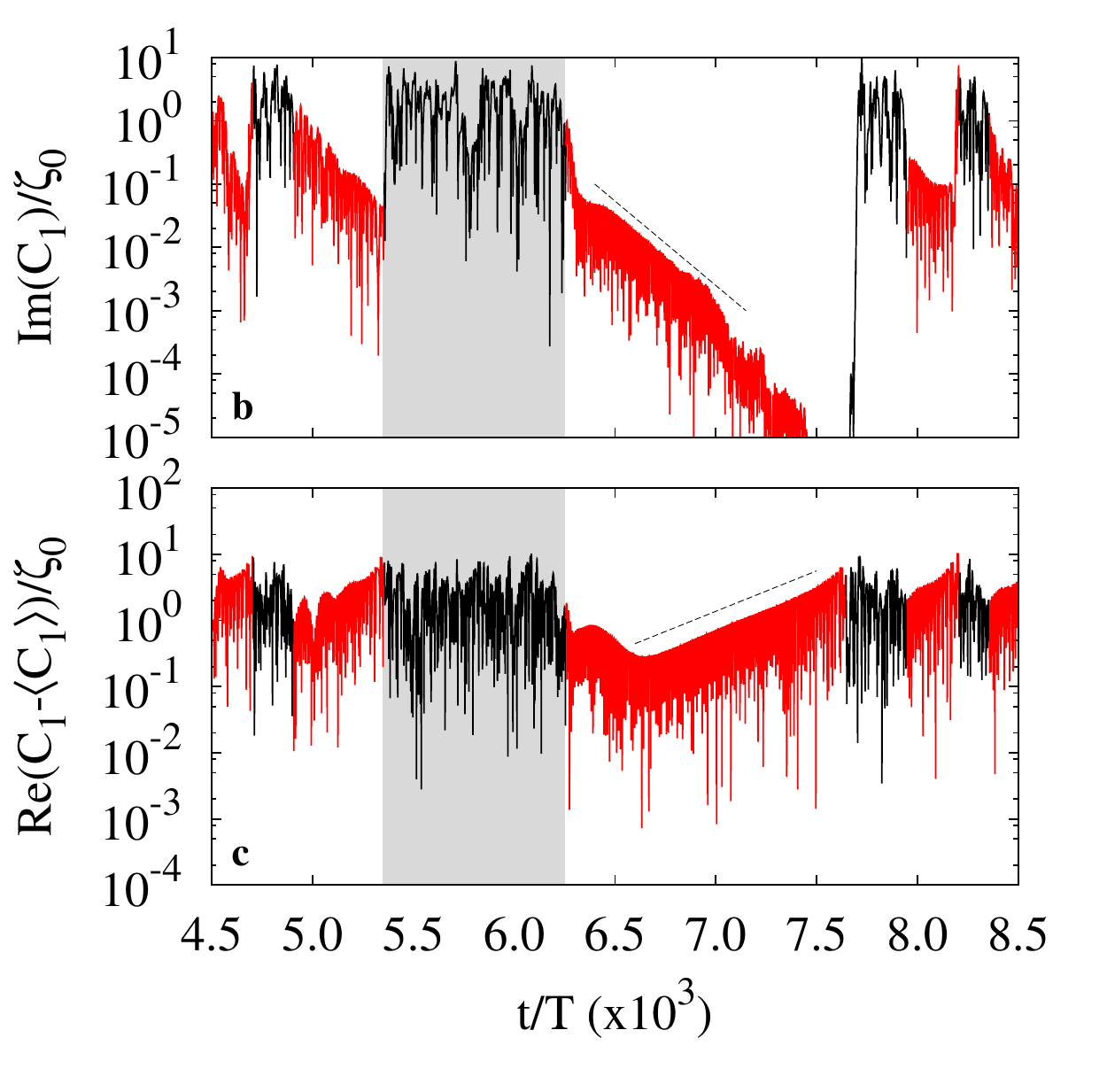}
	\caption{(color online) Instability characterization (a) 3D representation in the $(V, \mathrm{Re}(C_1)/\zeta_0,\mathrm{Im}(C_1)/\zeta_0)$-space for a trajectory at $\mathrm{Me} = 250$ (vicinity of the grey part Figs~\ref{fig:Fig2} (b,c). The slow outward spiraling flow of the laminar phase alternates with fast out-of-phane inward injection of the erratic phase, signature of a Shil'nikov type chaos. (b,c) Time series of $\mathrm{Im}(C_1)/\zeta_0$ and its centered quadrature $\mathrm{Re}(C_1-\langle C_1\rangle)/\zeta_0$ in lin-log scale.  Dashed lines evidence exponential evolution and  provide the convergence $t_I$ and divergence $t_D$ time. The color code is as in Fig.~\ref{fig:Fig2} (b,c).}
	\label{fig:Fig3}
\end{figure}

To analyse further the chaotic nature of the wave dynamics, we decompose the wave field using Graf's theorem into the co-moving Frenet basis centred at the position of the particle (see Supplemental Materials). The wave force is decomposed into a tangential component $\mathrm{Re}(C_1)$ and a normal component $\mathrm{Im}(C_1)$. A subset of the flow for $\mathrm{Me} = 250$ is plotted in Fig.~\ref{fig:Fig3}(a), in the three dimensional phase space $(V, \mathrm{Re}(C_1)/\zeta_0,\mathrm{Im}(C_1)/\zeta_0)$. We observe a converging flow along the axis $\mathrm{Im}(C_1)/\zeta_0$ followed by a diverging spiral in the plane $(V, \mathrm{Re}(C_1)/\zeta_0)$. This type of trajectory in phase space is encountered in Shil'nikov type chaos in which the flow converges toward a saddle point before diverging by spiralling outward in a plane. To evidence the nature of this chaos, we compute the saddle index $\nu = t_I/t_D$ \cite{Silva1993} which measures the ratio between the reinjection time scale $t_I$ and the time scale $t_D$ of the diverging flow. They are measured from the temporal evolution of $\mathrm{Im}(C_1)/\zeta_0$ and its quadrature $\mathrm{Re}(C_1-\langle C_1\rangle)/\zeta_0$ in the vicinity of the saddle (see Figs~\ref{fig:Fig3}(b) and~\ref{fig:Fig3}(c)). We find $\nu = 0.05$ which is much smaller than 1 and necessary for the existence of a Shil'nikov chaos~\cite{Silva1993}. Shil'nikov chaos has been observed in several contexts from Belousov-Zhabotinsky reaction~\cite{Argoul_1987}, electrode dissolution~\cite{Noh_2009,Bassett_1988}, Chua oscillator~\cite{Chakraborty_2010} to CO2 laser~\cite{Arecchi_1987,Dangoisse_1987}. It is encountered when an homoclinic cycle interacts with a subcritical Hopf bifurcation~\cite{Richetti_1986}. Exponential distribution of the laminar phase have been demonstrated in the context of type-II intermittency~\cite{Koronovskii_2008}. We find here that Shil'nikov chaos exhibits the same feature. 

\begin{figure}[tbhp]
	\centering
	\includegraphics[width=0.9\linewidth]{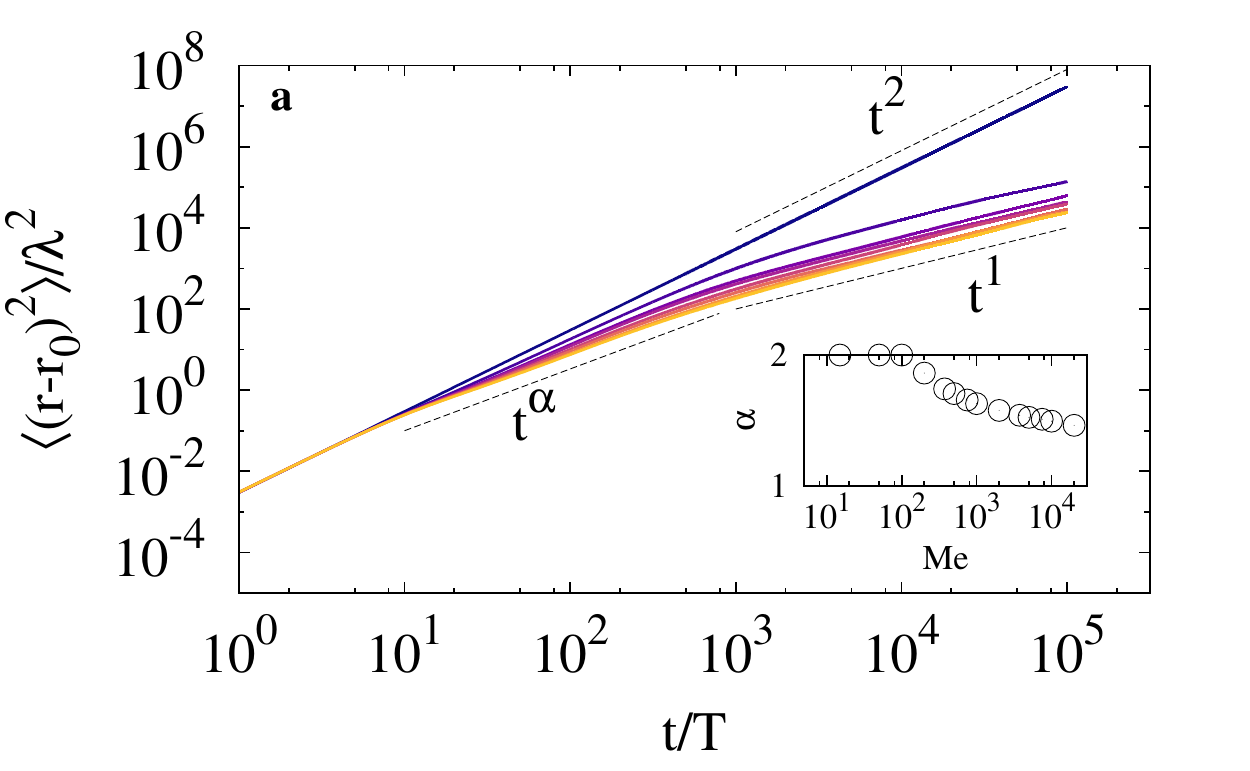}
	\includegraphics[width=0.9\linewidth]{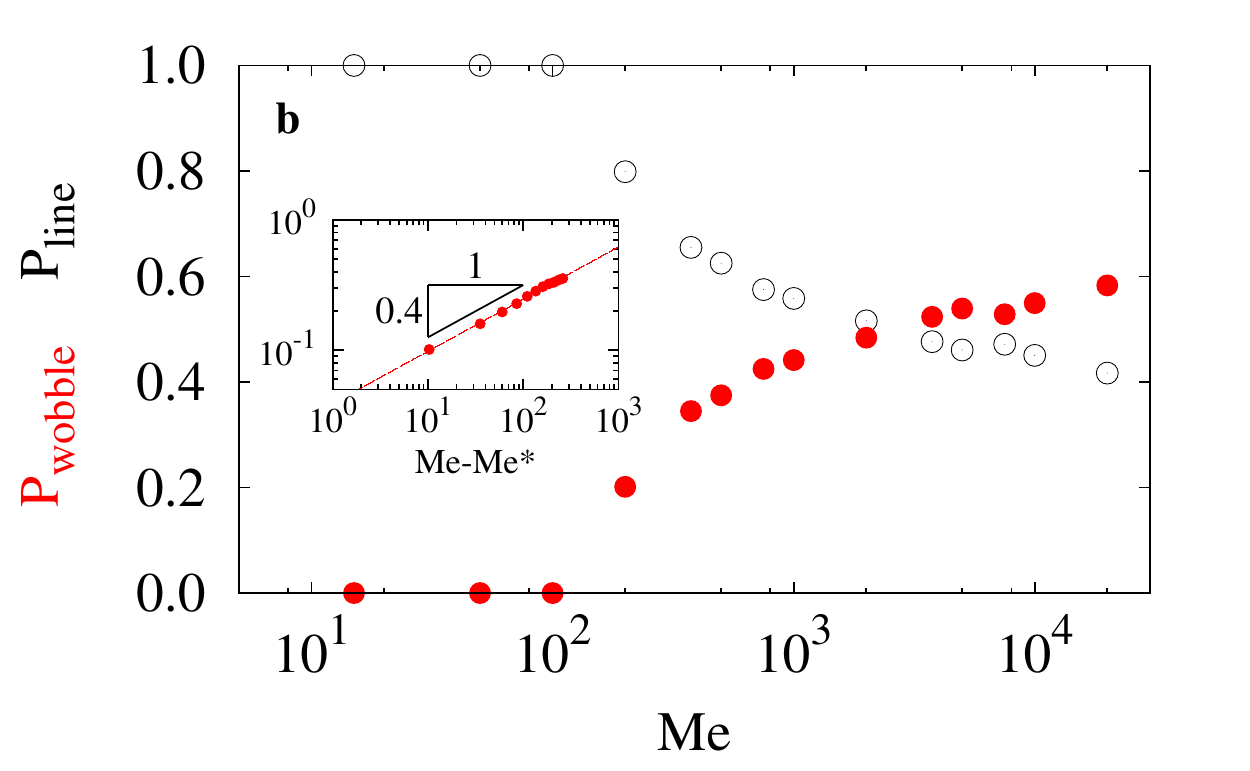}
	\caption{(color online). Statistical properties of long-time dynamics (a) MSD for increasing values of the memory parameter $\mathrm{Me}$ (From blue to yellow: $100, 200, 375, 500, 1000, 2000, 5000, 10^4$) as a function of the time $\Delta t$. Three regions can be distinguished: ballistic ($\frac{\Delta t}{T} < 10 $), superdiffusive ($10 < \frac{\Delta t}{T} < 10^3 $) and diffusive ($\frac{\Delta t}{T} > 10^3$). (Inset) Evolution of $\alpha$ with $\mathrm{Me}$ in lin-log. scale. (c) Measure of $\mathrm{P}_{\mathrm{wobble}}$ (resp. $\mathrm{P}_{\mathrm{line}}$) the proportion of time spent in the tumble phase (Black circle) (resp. in the running phase (red circle)). Inset, $\mathrm{P}_{\mathrm{wobble}}$ presents a scaling law $ \mathrm{P}_{\mathrm{wobble}} \sim (\mathrm{Me} - \mathrm{Me}^\star)^{0.4}$.}
	\label{fig:Fig4}
\end{figure}

Finally, we analyze the statistical properties of the long term dynamics by measuring the normalized Mean Squared Displacement (MSD) of the particle $\langle \vert\vec{r}-\vec{r}_0 \vert^2\rangle$. Figure~\ref{fig:Fig4}(a) shows the MSD as a function of time $\Delta T$ for increasing memory parameters and indicates several regimes. For $\mathrm{Me} = 100$, the MSD scales as $\Delta t^2$, as expected for a ballistic motion. For larger memory, the dynamics exhibit three regimes. At short time $\Delta t/T < 10$, a MSD typical of ballistic motion is recovered. At intermediate time scale $10 < \Delta t/T < 10^3$, we observe super-diffusive motion, in which the local MSD exponent $\alpha$ defines as $\alpha = d(\log(\langle \vert\vec{r}-\vec{r}_0 \vert^2\rangle)/d(\Delta T)$ lies in the range $ [1,2]$. The evolution of $\alpha$ as a function of the memory parameter $\mathrm{Me}$ is indicated in the inset of Fig.~\ref{fig:Fig4}(b). Its value is a direct consequence of the proportion of time spent in both phases of motion and we measure $\mathrm{P}_{\mathrm{line}}$  and $\mathrm{P}_{\mathrm{wobble}}$ as the fraction of time spend in linear and erratic phase respectively. Figure~\ref{fig:Fig4}(c) shows the evolution of $\mathrm{P}_{\mathrm{line}}$ and $\mathrm{P}_{\mathrm{wobble}}$ with $\mathrm{Me}$. We identify a critical memory parameter $\mathrm{Me}^\star = 140 \pm 5$ corresponding to the onset of a bimodal dynamics, above which a Shil'nikov chaos is triggered. For $\mathrm{Me} > \mathrm{Me}^\star$, $\mathrm{P}_{\mathrm{wobble}}$ increases, following a scaling law $ \mathrm{P}_{\mathrm{wobble}} \sim (\mathrm{Me} - \mathrm{Me}^\star)^\beta$ with an exponent $\beta  = 0.4 \pm 0.08$. In contrast the transition from the intermittent to the diffusive regime is progressive and we observe no signature of a discontinuity. 

In this Letter, we have studied the dynamics of a particle propelled by a self-generated wave field in two dimensions. The wave field creates a time-dependent erratic environment with long temporal coherence, coupled to the particle time evolution. The temporal damping of the waves controls a transition from a purely ballistic motion to erratic switches between ballistic and local diffusive motions. The characteristics of this purely deterministic dynamics are set by the properties of a Shil'nikov chaos occuring at the vicinity of a saddle point interacting with a Hopf bifurcation. The concomitance of ballistic, superdiffusive and diffusive dynamics for various time scales is usually encountered in the context of intermittent search strategies \cite{Benichou2011,Chupeau2015}. Especially, the dynamics illustrated in this article reminds the \emph{run and tumble} dynamics of chemotactic bacteria \cite{Berg1972}. In our case of a purely deterministic dynamics, it is interesting to note that such a multiscale feature can be encoded here by a simple deterministic wave-memory. It is worth noticing that this numerical experiment does not involve external noises neither interactions with other particles so that the diffusive behavior of the particle results from the wave-memory only. 

\begin{acknowledgments}
The authors thank V. Bacot and E. Fort for insightful discussions. This work was financially supported by the Actions de Recherches Concert\'{e}es (ARC) of the Belgium Wallonia-Brussels Federation under Contract No. 12-17/02. M.L., S.P and Y.C. acknowledge the financial support of the French Agence Nationale de la Recherche, through the project ANR Freeflow ANR-11-BS04-0001. Computational resources have been provided by the Consortium des \'{E}quipements de Calcul Intensif (C\'{E}CI), funded by the Fonds de la Recherche Scientifique de Belgique (F.R.S.-FNRS) Grant No. 2.5020.11.
\end{acknowledgments}

\bibliography{pnas-sample}

\end{document}